\documentclass[conference]{IEEEtran}
\IEEEoverridecommandlockouts
\usepackage{multirow}
\usepackage{graphicx}
\usepackage{booktabs}
\usepackage{subcaption}
\usepackage{float}
\usepackage{enumitem}
\usepackage{balance}

\usepackage{subcaption}
\usepackage{adjustbox}
\usepackage{caption}
\usepackage{url}
\newcommand{\stitle}[1]{\noindent\textup{\textbf{#1}}}  
\newcommand{\ourmodel}{$\mathsf{INTERPOS}$}
\newcommand{\mbr}{\ensuremath{\mathit{\mathsf{MobileRec}}}}
\newcommand{\mbraction}{\ensuremath{\mathit{\mathsf{Action}}}}
\newcommand{\mbrrole}{\ensuremath{\mathit{\mathsf{RolePlaying}}}}
\newcommand{\mbrpuzzle}{\ensuremath{\mathit{\mathsf{Puzzle}}}}
\newcommand{\mbrcasual}{\ensuremath{\mathit{\mathsf{Casual}}}}
\newcommand{\mbrsimulation}{\ensuremath{\mathit{\mathsf{Simulation}}}}
\newcommand{\mbrstrategy}{\ensuremath{\mathit{\mathsf{Strategy}}}}
\usepackage{cite}
\usepackage{amsmath,amssymb,amsfonts}
\usepackage{algorithmic}
\usepackage{graphicx}
\usepackage{textcomp}
\usepackage{xcolor}
\def\BibTeX{{\rm B\kern-.05em{\sc i\kern-.025em b}\kern-.08em
    T\kern-.1667em\lower.7ex\hbox{E}\kern-.125emX}}
\begin{document}

\title{INTERPOS: Interaction Rhythm Guided Positional Morphing for Mobile App Recommender Systems}
%\\
%{\footnotesize \textsuperscript{*}Note: Sub-titles are not captured for https://ieeexplore.ieee.org  and should not be used}

\author{\IEEEauthorblockN{M.H. Maqbool$^\clubsuit$, Moghis Fereidouni$^\heartsuit$, Umar Farooq$^\spadesuit$, A.B. Siddique$^\heartsuit$, Hassan Foroosh$^\diamondsuit$}
\IEEEauthorblockA{
hasan.ucf@gmail.com$^\clubsuit$, moghis.fereidouni@uky.edu$^\heartsuit$, ufarooq@lsu.edu$^\spadesuit$, siddique@cs.uky.edu$^\heartsuit$, hassan.foroosh@ucf.edu$^\diamondsuit$ \\
\textit{Independent Researcher$^\clubsuit$, University of Kentucky$^\heartsuit$, Louisiana State University$^\spadesuit$, University of Central Florida$^\diamondsuit$} 
}
}
% \and

\maketitle
\begin{abstract}
%Recommender systems are pervasive in digital platforms, from suggesting products on online e-commerce stores to recommending TV shows and movies on streaming platforms. 
% given the wide range of products offered to users, 
The mobile app market has expanded exponentially, offering millions of apps with diverse functionalities, yet research in mobile app recommendation remains limited.
Traditional sequential recommender systems utilize the order of items in users' historical interactions to predict the next item for the users. 
Position embeddings, well-established in transformer-based architectures for natural language processing tasks, effectively distinguish token positions in sequences.
In sequential recommendation systems, position embeddings can capture the order of items in a user's historical interaction sequence. 
Nevertheless, this ordering does not consider the time elapsed between two interactions of the same user (e.g., 1 day, 1 week, 1 month), referred to as ``user rhythm''. 
In mobile app recommendation datasets, the time between consecutive user interactions is notably longer compared to other domains like movies, posing significant challenges for sequential recommender systems.
To address this phenomenon in the mobile app domain, we introduce {\ourmodel}, an \underline{Inter}action Rhythm Guided \underline{Pos}itional Morphing strategy for autoregressive mobile app recommender systems. 
{\ourmodel} incorporates rhythm-guided position embeddings, providing a more comprehensive representation that considers both the sequential order of interactions and the temporal gaps between them.
This approach enables a deep understanding of users' rhythms at a fine-grained level, capturing the intricacies of their interaction patterns over time.
%By integrating user rhythm into positional embeddings, we aim to enhance the system's ability to discern not only the order but also the timing of user preferences.
%We present {\ourmodel}, an \underline{Inter}action Rhythm Guided \underline{Pos}itional Morphing strategy, which incorporates rhythm-guided position embeddings for recommender systems to reflect the global view, by effectively capturing user's preferences.
% Can interaction rhythm-guided position embeddings for recommender systems reflect the global view, by effectively capturing user's preferences? 
%We empirically establish that the user's interaction rhythm strongly correlates with the user's preferences. 
We propose three strategies to incorporate the morphed positional embeddings in two transformer-based sequential recommendation system architectures. 
Our extensive evaluations show that {\ourmodel} outperforms state-of-the-art models using 7 mobile app recommendation datasets on NDCG@K and HIT@K metrics.
%Our extensive evaluations show that {\ourmodel} outperforms state-of-the-art models, across all the categorical splits of the {\mbr} dataset.
%Notably, there is an average improvement of 73.23\% and 69.24\% in NDCG@5 and NDCG@10 metrics, respectively. 
%In HIT@5 and HIT@10, we report an improvement of 70.5\% and 65.61\%.  
The source code of {\ourmodel} is available at \url{https://github.com/dlgrad/INTERPOS}.

\end{abstract}


\section{Introduction}
\begin{figure}
    \centering
    %\vspace{20pt}
    %\includegraphics[width=\linewidth]{src/images/introduction/interpos_intro.pdf}
    \includegraphics[width=0.96\linewidth]{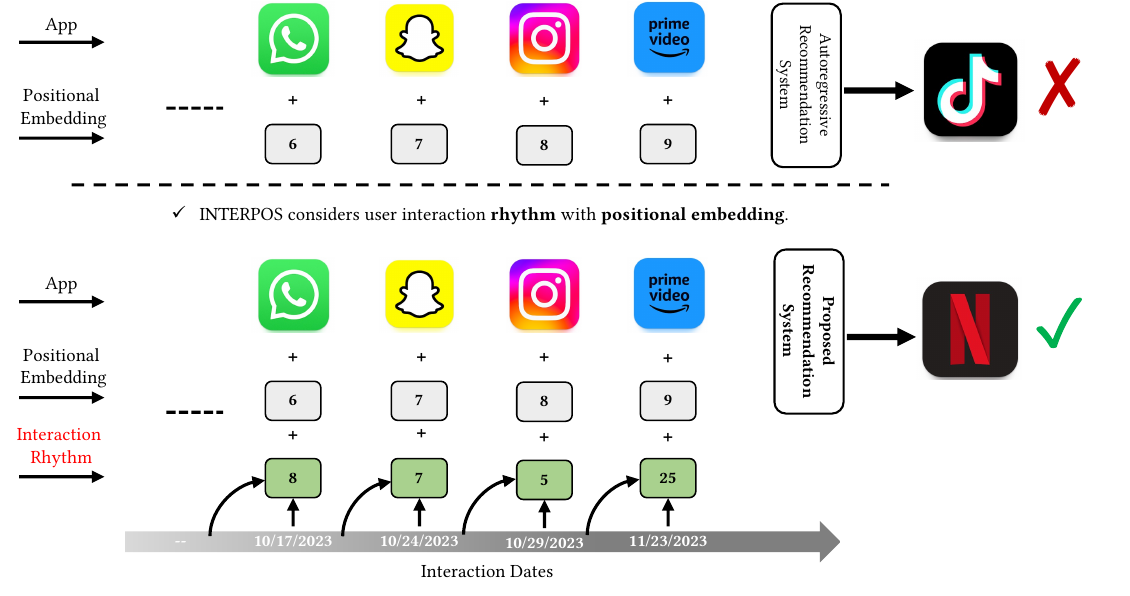}
    \vspace{-5pt}
    \caption{INTERPOS considers user interaction rhythm with position embedding for tracking user preferences effectively reflecting the user's behavior over time.
}
\label{fig:interpos-intro}
\vspace{-15pt}
\end{figure}
%With the exponential rise of mobile platforms, the mobile app market offers an unprecedented collection of applications offering diverse functionalities. 
Mobile app market is witnessing exponential growth.
Apple Appstore~\cite{appstore} and Google Play~\cite{googleplay} include over $2.2$ and $3.5$ million apps, respectively~\cite{number-of-apps-on-stores}.
%The scale of selection at the user's disposal poses a significant challenge to users to discover their preferred apps. 
%Recommender systems play a crucial role in alleviating users' cognitive load and streamlining the product search process to align with individual preferences. 
%A recommender system garners the utmost importance to reduce the user's cognitive strain by filtering and recommending apps aligned with the user's preferences. 
Despite the vastness of the market and the abundance of apps accessible to users, the research in mobile app recommendation remains constrained.
%Irrespective of the size of the market and the number of apps available to users, the research in the mobile app reco,m,endation remains constrained. 
%Figure~\ref{fig:interpos-intro} shows a typical scenario, where recommendation systems can suggest new apps to users based on their previous interactions and installations.  

% In recent times, with considerable advances in technology, the consumer market has witnessed rapidly growing product lines. 
% From household utility appliances to grocery items, home improvement to beauty and personal care products, general merchandise to specialized equipment, DIY hardware tools to consumer electronics, and from clothing to footwear, a consumer is faced with a huge collection of products to choose from. 
% Correspondingly, the mobile app market offers an unprecedented collection of applications offering a diverse range of functionality and services. 
% For example, Google Play serves around 2.5 billion users around the world and offers millions of apps belonging to a wide range of categories. 
% Keeping in view the scale of selection at the user's disposal, a recommender system garners the utmost importance to reduce the user's cognitive strain by filtering and recommending the apps that are aligned with the user's preferences. 

Sequential recommender systems play a crucial role by leveraging the temporal order of items in users' historical interactions to predict their future preferences. 
In natural language processing tasks, position embedding is a well-established technique in transformer-based architectures to capture the order of tokens in sequences. 
However, in sequential recommendation systems, the reliance on position embeddings falls short in addressing a critical aspect -- the temporal dynamics inherent in users' interaction patterns, which we refer to as ``user rhythm''. 
While position embedding can distinguish the order of items in a user's historical interactions, it does not take into account the time elapsed between two interactions by the same user, be it a day, a week, or a month.
In mobile app recommendation datasets, the time gap between two consecutive user interactions is considerably longer compared to other widely studied domains, such as movies.

\begin{figure}
    \centering
    \includegraphics[scale=0.34]{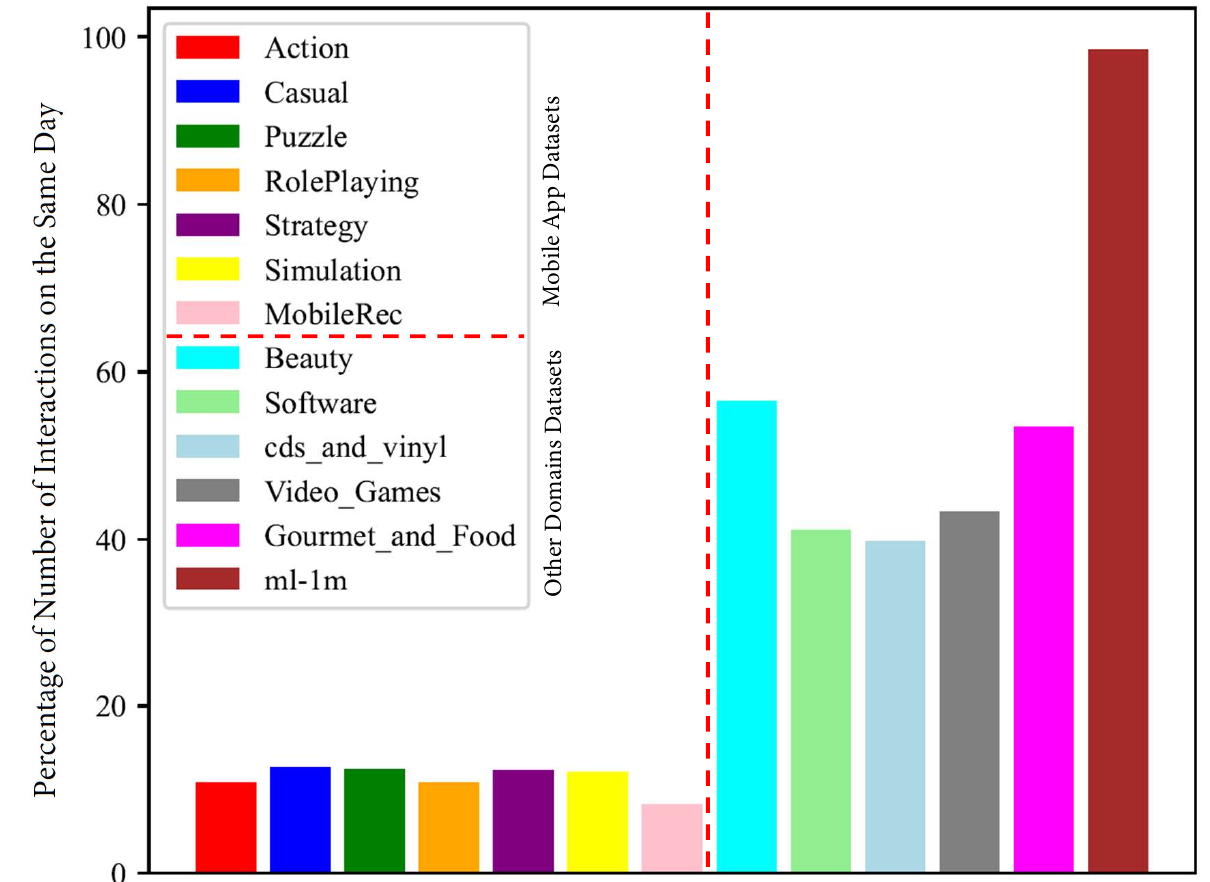}
    \caption{Percentage of consecutive user interactions on the same day in the mobile app domain compared to other domains.}
    \label{fig:rhythm_emergence_trends}
    \vspace{-14pt}
\end{figure}

To quantify this phenomenon, we analyze the time intervals between consecutive user interactions in mobile app recommendation datasets and compare them with other domains, including various categories from the \textsc{Amazon Product Reviews} dataset \cite{mcauley2015image, he2016ups, ni2019justifying} (e.g., Beauty, Video Games, CDs and Vinyl, Software, Grocery and Gourmet Food) and \textsc{MovieLens-1M} \cite{harper2015movielens}.
Our analysis, detailed in Figure~\ref{fig:rhythm_emergence_trends}, reveals an important pattern in other domain datasets: a substantial concentration of zero time differences between user interactions. 
Since many datasets, except for mobile app recommendation datasets, exhibit interactions on the same day, existing methods~\cite{li2020time,du2023frequency} that consider time intervals between interactions do not need to learn to account for extended time gaps.
These methods employ interval-based self-attention mechanisms, which are not effective in capturing the longer gaps between interactions.

%This challenge persists even for existing systems~\cite{li2020time,du2023frequency} that already take into account time intervals between user interactions.
%This prolonged period between interactions presents a unique challenge for developing effective sequential mobile app recommender systems even for existing sequential recommender systems that consider time intervals between user interactions~\cite{li2020time,du2023frequency}. 
%
%This omission becomes a significant limitation as it neglects to incorporate the evolving patterns over time in user rhythm, which has been empirically shown to exhibit a strong correlation with user preferences. 

In this work, we introduce a novel approach that integrates the gaps between successive interactions directly at the embedding layer.
By embedding the time intervals early in the model architecture, we allow the model to learn and incorporate the temporal dynamics of user behavior more effectively. 
This approach captures longer gaps between interactions effectively, which are particularly prevalent in the mobile app domain, and significantly enhances the model's ability to generate accurate recommendations compared to existing techniques~\cite{li2020time,du2023frequency} that fail to adequately account for these extended time intervals.
Figure~\ref{fig:interpos-intro} provides an overview of our approach.
%Given the concentration of interactions on the same day in many datasets except mobile app recommendation datasets, existing methods~\cite{li2020time,du2023frequency} that apparently consider time intervals between user interactions can get away without doing too much.
%Specifically, these methods~\cite{li2020time,du2023frequency} propose self-attention mechanisms to consider time intervals between user interactions, which we argue can not effectively capture longer time intervals. 
%In this work, we propose to incorporate the interaction gaps between successive interactions early in the model significantly improves the recommendations.  
%Figure~\ref{fig:interpos-intro} provides an overview of our approach.
%In this work, we argue that incorporating interaction gaps (we call it \textbf{user's interaction rhythm}) between successive user/item interactions improves the recommendation systems.  

Specifically, we build {\ourmodel} on the observation that a user's interaction rhythm encodes valuable behavioral patterns that can be used for morphing the position encoding of items effectively.
%we propose to morph the positioning with the user's interaction rhythm, which allows the model to learn the user's preference biases and corresponding shifts in the user's choices. 
To achieve this, we introduce three fusion-based strategies for transformer-based recommender system architectures. 
These strategies, namely basic fusion, multilayer perceptron-based fusion, and gated fusion, are developed to integrate user rhythm into the recommendation process.
The basic fusion strategy adds traditional position embeddings with rhythmic embeddings. 
The multilayer perceptron-based fusion incorporates additional layers of multilayer perceptrons after concatenating absolute position and user rhythm embeddings, enhancing the model's capacity to capture intricate relationships. 
Lastly, the gated fusion leverages a gating mechanism, enabling the model to learn how to judiciously mix and match positional embedding with rhythmic embedding.

To validate the efficacy of our proposed strategies, we implement them in two well-established transformer-based architectures: LightSANs~\cite{fan2021lighter} and SASRec~\cite{kang2018self}. 
Through this integration, we report considerable performance improvements across 7 mobile app recommendation datasets.
On average, across all the datasets, {\ourmodel} fusion strategies, called INTERPOS-BF, INTERPOS-GF, and INTERPOS-MF, achieve up to 157.3\%, 156.67\%, and 158\% 
% 73.16\%, 71.87\%, and 73.17\% 
improvement in \textit{NDCG@20} and 145.62\%, 141.01\%, and 142.85\%
% 45.15\%, 66.45\%, and 69.72\% 
improvement in \textit{NDCG@20}, respectively. 
We observe that, across all the dataset splits, INTERPOS-BF, INTERPOS-GF, and INTERPOS-MF show up to 145.15\%, 139.39\%, and 1143.63\% improvement in \textit{HIT@10} and 142.6\%, 133.04\%, and 137.04\% improvement in \textit{HIT@20}.

Our contributions are as follows:

\begin{itemize}
\item We investigate an underexplored domain in recommendation systems: mobile app recommendations.

\item We introduce a novel paradigm that seamlessly integrates position embedding with user interaction rhythm by developing three fusion strategies for transformer-based recommender system architectures.

\item We empirically demonstrate the effectiveness of incorporating interaction rhythm guidance and morphed positioning through extensive studies in mobile app recommendation, showcasing its superior performance.
\end{itemize}

% \begin{figure*}
%   \centering
% \includegraphics[scale=0.4]{src/images/introduction/intro_digram_cropped.pdf}
    
%   \caption{This is the introduction diagram.}
% \label{fig:intro_diagram}
% \end{figure*}

% \begin{figure*}
%   \centering

% \includegraphics[width=\textwidth]{src/images/introduction/inter_rhythm_heatmap_plot_cropped.pdf}
    
%   \caption{User interaction rhythm}
%   \label{fig:rhythm_heatmap}
% \end{figure*}

%RQs:
%1. Does incorporating rhythm help models perform better?
%--> Show all subsets of the dataset + full dataset results. The underlying model is LightSANS.

%2.  What is the best way to incorporate rhythm?
%--> Discuss what is/are better than others and when.

%3. Does incorporating rhythm in any underlying models help to get better results?
%SASRec.
%--> 3 approaches show better results?

%Experimental details:
%one large-scale dataset; 6 subsets.
%3-6 rhythm incorporation approaches;
%2 underlying models 

%ufarooq
%1. We show the benefit of using rhythm.
%2. We propose 3 strategies to incorporate rhythm.
%3. Our experiments show after incorporating 2 architectures, rhythm works and outperforms. 
\section{Preliminaries}

\begin{figure*}
  \centering
  \begin{subfigure}{0.3\textwidth}
    \hspace{-0.5cm}
    \includegraphics[scale=0.55]{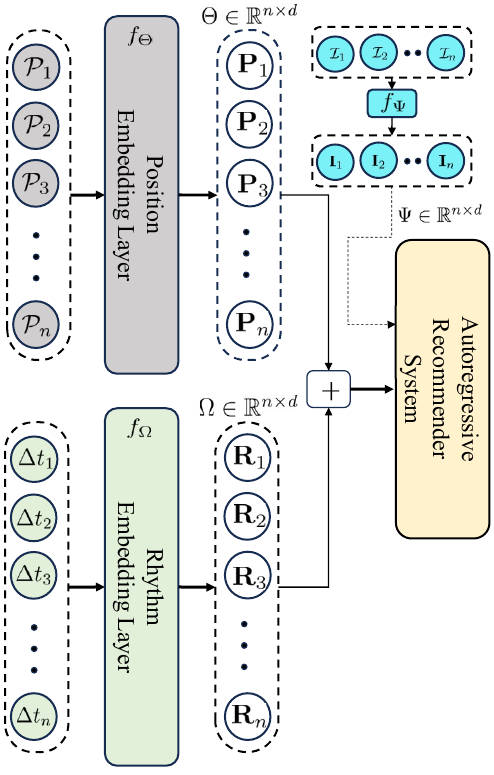}
    \caption{INTERPOS-BF}
    \label{fig:interposbf}
  \end{subfigure}
  \hspace{0.01\textwidth}
  \begin{subfigure}{0.3\textwidth}
  \hspace{-0.7cm}
    \includegraphics[scale=0.55]{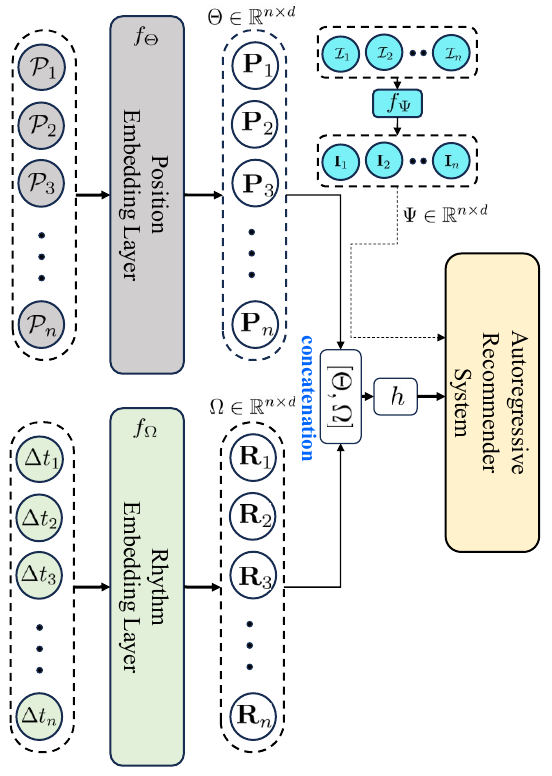}
    \caption{INTERPOS-MF}
    \label{fig:interposmf}
  \end{subfigure}
  \hspace{0.01\textwidth}
  \begin{subfigure}{0.33\textwidth}
    \hspace{-0.7cm}
    \includegraphics[scale=0.55]{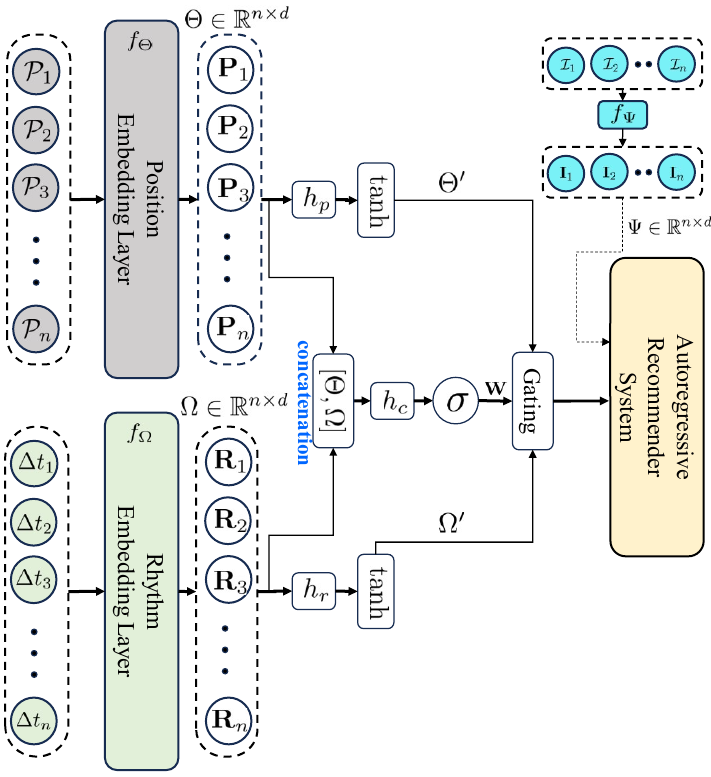}
    \caption{INTERPOS-GF}
    \label{fig:interposgf}
  \end{subfigure}

  \caption{Overview of {\ourmodel} Fusion Architectures.}
  \label{fig:interpos_model_diagrams}
  \vspace{-12pt}
\end{figure*}
%\footref{interpos_all}

\stitle{Problem Formulation.}
Let $\mathcal{U} = \left \{ \mathcal{U}_1, \mathcal{U}_2,\mathcal{U}_3,\dots,\mathcal{U}_{\left | \mathcal{U} \right |}  \right \}$ and $\mathcal{I} = \left \{ \mathcal{I}_1, \mathcal{I}_2, \mathcal{I}_3, \dots,\mathcal{I}_{\left | \mathcal{I} \right |} \right \}$ represent the set of users and apps, respectively.
A user-app interaction sequence is defined by $\mathcal{S}^{\mathcal{U}_{i}} = \left \{ \left ( \mathcal{U}_i, \mathcal{I}_p \right )_{|\mathcal{S}^{\mathcal{U}_i}|} \right \}$ where  
$i \in \left \{ x \mid 1 \leq x \leq \mathcal{U}_{\left | U \right |}  \right \}
$ and $p \in \left \{ y \mid 1 \leq y \leq \mathcal{I}_{\left | I \right |}  \right \}$.
Given a user-app interaction sequence for $i$-th user: 
\vspace{-4pt}
\begin{equation}
\mathcal{S}^{{\mathcal{U}}_i}=\left \{\left ( \mathcal{U}_i, \mathcal{I}_1 \right )_1, \left ( \mathcal{U}_i, \mathcal{I}_2 \right )_2, \dots, \left ( \mathcal{U}_i, \mathcal{I}_p \right )_{|\mathcal{S}^{\mathcal{U}_i}|-1} \right \}
\label{eq:ui_seq}
\end{equation}
up to sequence length  $|\mathcal{S}^{\mathcal{U}_i}|-1$,  sequential recommendation seeks to predict the next app which a user is likely to interact with, $\mathcal{I}_{p+1}$ in the user's next interaction $\left ( \mathcal{U}_i, \mathcal{I}_{p+1} \right )_{|\mathcal{S}^{\mathcal{U}_i}|}$. 
It is important to note that the users are not explicitly modeled in the task formulation. 
The users are represented by their interaction history. 
Having noted this detail, we can simplify the notation by dropping the reference to the user in the interaction history, $\mathcal{S}^{{\mathcal{U}}_i}$. 
The simplified user-app interaction sequence can be written as: 
\vspace{-4pt}
\begin{equation}
    \mathcal{S}^{{\mathcal{U}}_i}=\left ( \mathcal{S}_1, \mathcal{S}_2, \mathcal{S}_3, \mathcal{S}_4, \dots, \mathcal{S}_{|\mathcal{S}^{{\mathcal{U}}_i}|}  \right )
    \label{eq:ui_seq_short}
\end{equation}
\vspace{-16pt}

Given an input user/item sequence  $\mathcal{S}^{{\mathcal{U}}_i}=\left ( \mathcal{S}_1, \mathcal{S}_2, \mathcal{S}_3, \mathcal{S}_4, \dots, \mathcal{S}_{|\mathcal{S}^{{\mathcal{U}}_i}|-1}  \right )
$
the model is expected to predict the shifted version of the input
$\mathcal{S}^{{\mathcal{U}}_i}=\left ( \mathcal{S}_2, \mathcal{S}_3, \mathcal{S}_4, \mathcal{S}_5, \dots, \mathcal{S}_{|\mathcal{S}^{{\mathcal{U}}_i}|}  \right )$. 
Sequential data can be modeled with an autoregressive transformers architecture~\cite{vaswani2017attention,radford2018improving, roberts2019exploring, devlin2018bert, yang2019xlnet, keskar2019ctrl, dong2019unified, liu2019roberta, he2020deberta}. %Transformer~\cite{} lies central to the autoregressive modeling of sequential data.

\stitle{Self-attention.}
Self-attention is the main architectural choice behind transformers. 
Self-attention is defined as:
$
\operatorname{Attention}(Q, K, V)=\operatorname{softmax}\left(\frac{Q K^T}{\sqrt{d_k}}\right) V,
$ where $Q, K$, and $V$ are query, key, and value matrices, respectively. 
In a transformer-based next-item prediction task, an embedding layer, $f_{\Psi}$, is employed to encode the items in a user's interaction history, $\mathcal{S}^{{\mathcal{U}}_i}$, into the embedding space. 
\vspace{-4pt}
\begin{equation}
    \Psi = f_{\Psi}(\mathcal{S}^{{\mathcal{U}}_i}),
\end{equation}
where item embedding matrix $\Psi \in \mathbb{R}^{n  \times d}$ and $d$ is the embedding dimension. 
Correspondingly, the input embedding matrix can be thought of as containing item embeddings. 

% and represented as:
% %shown in Equation~\ref{equ:input_emb_matrix}. 

% % \begin{equation}
% %     \Psi = \left [
% %     \mathbf{I}_1, 
% %     \mathbf{I}_{2}, 
% %     \dots,
% %     \mathbf{I}_{n}
% %                 \right ]^\top
% %     \label{equ:input_emb_matrix}
% % \end{equation}

% \begin{equation}
%     \Psi = \begin{bmatrix}
%     \mathbf{I}_1\\ 
%     \mathbf{I}_{2}\\ 
%     \vdots\\ 
%     \mathbf{I}_{n}
%                 \end{bmatrix}
%     \label{equ:input_emb_matrix}
% \end{equation}

Each $\mathbf{I}_i$ represents an item in the embedding space. 
The self-attention mechanism must be equipped with a sense of order in the input sequence. 
Let us define the absolute positions as $\mathbf{P}=\left\{\mathbf{P}_1, \mathbf{P}_1, \dots, \mathbf{P}_n \right\}$. 
$f_{\Theta}(.)$ is employed for encoding the absolute position information into embedding space, represented as $\Theta$. 
Position embeddings have the same dimension as the input embeddings $\Theta \in \mathbb{R}^{n \times d}$. Finally, $E = \Psi + \Theta$ represents the user-item interaction sequence, $\mathcal{S}^{{\mathcal{U}}_i}$ in the embedding space:%and can be represented as:

% % \begin{equation}
% %     \Theta = \left [
% %     \mathbf{P}_1, 
% %     \mathbf{P}_{2}, 
% %     \vdots,
% %     \mathbf{P}_{n}
% %                 \right ]
% %     \label{equ:pos_emb_matrix}
% % \end{equation}

% \begin{equation}
%     \Theta = \begin{bmatrix}
%     \mathbf{P}_1 \\
%     \mathbf{P}_{2} \\ 
%     \vdots, \\
%     \mathbf{P}_{n}
%                 \end{bmatrix}
%     \label{equ:pos_emb_matrix}
% \end{equation}

% \begin{equation}
%     \mathbf{E} = \left [
%     \mathbf{I}_1 + \mathbf{P}_1 \newline
%     \mathbf{I}_2+ \mathbf{P}_2 \\ 
%     \vdots \\
%     \mathbf{I}_n + \mathbf{P}_n
%                 \right ]
%     \label{equ:3}
% \end{equation}
\vspace{-4pt}
\begin{equation}
    \mathbf{E} = \begin{bmatrix}
    \mathbf{I}_1 + \mathbf{P}_1\\ 
    \mathbf{I}_2+ \mathbf{P}_2\\ 
    \vdots\\ 
    \mathbf{I}_n + \mathbf{P}_n
                \end{bmatrix}
    \label{equ:3}
\end{equation}

\section{Approach}
First, we conceptualize and represent the mathematical formulation of our proposed approach. 

\subsection{User's Interaction Rhythm}
A user-item interaction is represented by a sequence $\mathcal{S}^{{\mathcal{U}}_i}$ for a user $\mathcal{U}_i$. We assume that a user's interaction sequence emerges across a particular period, and we call it the \textit{Activity Window}. Referring back to Equation \ref{eq:ui_seq}, we can rewrite the Equation \ref{eq:ui_seq_short} as follows:
\vspace{-4pt}
\begin{equation}
    \mathcal{S}^{{\mathcal{U}}_i}=\left ( \mathcal{S}_1^{t_1}, \mathcal{S}_2^{t_2}, \mathcal{S}_3^{t_3}, \mathcal{S}_4^{t_4}, \dots, \mathcal{S}_{|\mathcal{S}^{{\mathcal{U}}_i}|}^{t_N}  \right )
    \label{eq:ui_seq_with_rhythm}
\end{equation}

where $t_N$ represents the time of interaction for the last interaction for a user $\mathcal{U}_i$. 
Let us define $\Delta t_1 = 0$ and $\Delta t_i = t_{i} - t_{i-1} \forall i \in \left\{2, 3, \dots, n\right\}$. Having the notion of $\Delta t_i$ noted, let us define $f_\Omega(.)$ to encode a user's interaction rhythm  $\Delta t_i$ to the embedding space. 
Now, we can define the user's embedded interaction rhythm, $\Omega$, as follows:

\vspace{-4pt}
\begin{equation}
    \Omega = \left \{ \mathbf{R}_1, \mathbf{R}_2, \mathbf{R}_3, \dots, \mathbf{R}_n \right \}
    \label{eq:interaction_rhythm}
\end{equation}

$\Omega$ encodes the useful behavioral and preferential shifts for a user over the \textit{Activity Window}. $\mathbf{R}_i$ represents the encoded $\Delta t_i$ in the embedding space, i.e., $f_{\Omega}(\Delta t_i)$.  
We argue that $\Omega$ can be consumed by next-item prediction architectures for making more informed predictions. We propose three different architectural variations to fuse $\Omega$ into an autoregressive next-item prediction architecture.
% \begin{figure*}
% \begin{minipage}[t]{0.3\textwidth}
%   \includegraphics[scale=0.5]{src/images/interposbf_model.pdf}
%   \caption{INTERPOS-BF}
%   \label{fig:interposbf}
% \end{minipage}%
% %\hfill % maximize the horizontal separation
% \begin{minipage}[t]{0.3\textwidth}
%   \includegraphics[scale=0.5]{src/images/interposmf_model.pdf}
%   \caption{INTERPOS-MF}
%   \label{fig:interposmf}
% \end{minipage}%
% % \hfill
% \begin{minipage}[t]{0.3\textwidth}
%   \includegraphics[scale=0.5]{src/images/interposgf_model.pdf}
%   \caption{INTERPOS-GF}
%   % \label{fig:interposgf}
% \end{minipage}%
% \end{figure*}

% \begin{figure*}
%   \centering

%   \begin{subfigure}{0.28\textwidth}
%     \includegraphics[scale=0.5
% ]{src/images/interposbf_model.pdf}
%     \caption{INTERPOS-BF}
%     \label{fig:sub1}
%   \end{subfigure}  
%   \begin{subfigure}{0.2\textwidth}
%     \includegraphics[scale=0.5]{src/images/interposmf_model.pdf}
%     \caption{INTERPOS-MF}
%     \label{fig:sub2}
%   \end{subfigure}
%   \begin{subfigure}{0.33\textwidth}
%     \includegraphics[scale=0.5
% ]{src/images/interposgf_model.pdf}
%     \caption{INTERPOS-GF}
%     \label{fig:sub3}
%   \end{subfigure}

%   \caption{Three Subfigures Spanning Two Columns}
%   \label{fig:subfigures}
% \end{figure*}

\begin{table*}[t]
\centering
\caption{Datasets statistics.}
\vspace{-5pt}
\begin{tabular}{l|ccccccc}
\toprule
\multicolumn{1}{c|}{\textbf{Descriptor} $\downarrow$ \textbf{Dataset} $\xrightarrow{}$}     & \multicolumn{1}{c}{\textbf{Action}} & \multicolumn{1}{c}{\textbf{RolePlaying}} & \multicolumn{1}{c}{\textbf{Casual}} & \multicolumn{1}{c}{\textbf{Simulation}} & \multicolumn{1}{c}{\textbf{Strategy}} & \multicolumn{1}{c}{\textbf{Puzzle}} & \multicolumn{1}{c}{\textbf{MobileRec}} \\ \hline
\textbf{\# Unique Users}                  & 80961                               & 77858                                    & 48091                               & 56771                                   & 52055                                 & 51456                               & 0.7 M                                  \\
\textbf{\# Unique Apps}                   & 529                                 & 658                                      & 432                                 & 537                                     & 520                                   & 537                                 & 10173                                  \\
\textbf{Avg. interactions per user}       & 9.82                                & 9.63                                     & 8.22                                & 8.58                                    & 8.37                                  & 8.34                                & 27.56                                  \\
\textbf{Avg. interactions per app}        & 1502.93                             & 1139.57                                  & 915.47                              & 906.56                                  & 838.06                                & 798.92                              & 1896.88                                \\
\textbf{Maximum interactions by a user}   & 29                                  & 30                                       & 23                                  & 24                                      & 22                                    & 24                                  & 256                                    \\
\textbf{Maximum interactions on an app}   & 5372                                & 3663                                     & 5209                                & 3180                                    & 305                                   & 3742                                & 14,345                                 \\
\textbf{Total Interactions (in Millions)} & 0.79 M                              & 0.74 M                                   & 0.39 M                              & 0.48 M                                  & 0.42 M                                & 0.43 M                              & 19.3 M                                \\
\bottomrule
\end{tabular}
\label{tbl_dataset_stats}
\vspace{-12pt}
\end{table*}

\subsection{Fusion Architectures}
We present three different fusion methodologies to incorporate $\Omega$ into autoregressive next-item prediction architectures. 
Our proposed methodologies morph the existing position encoding $\Theta$, by fusing $\Omega$ into it, effectively enriching the model's ability to discern the relative importance of the user's interaction history over the \textit{Activity Window}.

\stitle{Basic Fusion}. Basic fusion employs a direct fusion approach of $\Theta$ and $\Omega$. 
Let us define a vector-valued function $f: \mathbb{R}^d \times \mathbb{R}^d \to \mathbb{R}^d$. $f$ can represent element-wise summation, element-wise multiplication, or some other function. 
In our case, $F$ is a simple element-wise addition. 
Let us say $\mathbf{M}$ represents a morphed version of $\Theta$ by fusing it with $\Omega$ through $f$:

\vspace{-6pt}
\begin{equation}
    \mathbf{M} = f(\Theta, \Omega)
\end{equation}

$\mathbf{M}$ is employed to update Equation \ref{equ:3} to get the user-app interaction sequence. $\mathcal{S}^{{\mathcal{U}}_i}$ in the updated embedding space,  $\mathbf{E'}$, with morphed position embeddings,

\vspace{-6pt}
\begin{equation}
    \mathbf{E'} = \begin{bmatrix}
    \mathbf{I}_1 + \mathbf{M}_1 \\ 
    \mathbf{I}_2+ \mathbf{M}_2 \\
    \vdots \\
    \mathbf{I}_n + \mathbf{M}_n
                \end{bmatrix}
    \label{equ:bm}
\end{equation}

% \begin{equation}
%     \mathbf{E'} = \left [
%     \mathbf{I}_1 + \mathbf{M}_1, 
%     \mathbf{I}_2+ \mathbf{M}_2, 
%     \dots,
%     \mathbf{I}_n + \mathbf{M}_n
%                 \right ]^\top
%     \label{equ:bm}
% \end{equation}
where $\mathbf{M}_i$ is an element-wise sum of $\mathbf{P}_i$ and $\mathbf{R}_i$.
Figure~\ref{fig:interposbf} depicts the INTERPOS-BF architecture.

\stitle{MLP Fusion}. 
Let us define $h$ to be a multi-layer perceptron such that $h: \mathbb{R}^{2d} \to \mathbb{R}^d$. 
Given $\Theta$ and $\Omega$, $\left [\Theta, \Omega \right]$ represent their instance-wise concatenation, resulting in $\mathbb{R}^{n \times 2d}$ matrix. 
Having $\Theta$ and $\Omega$ concatenated, $h$ is employed to generate the fused matrix $\mathbf{M}$,

\vspace{-6pt}
\begin{equation}
    \mathbf{M} = h( \left[\Theta, \Omega \right])
\end{equation}

Equation \ref{equ:bm} is employed for generating the updated embeddings for $\mathcal{S}^{{\mathcal{U}}_i}$ where $\mathbf{M}_i$ represents $\mathbf{h}(\left[\mathbf{P}_i, \mathbf{R}_i\right])$ 
Figure~\ref{fig:interposmf} outlines the INTERPOS-MF fusion architecture.

\stitle{Gated Fusion.}
Let $h_p$, $h_r$ and $h_c$ be multi-layer perceptrons. $\Theta$ is linearly projected using $h_p$ and $
\tanh$ non-linearity is applied such that $h_p(\Theta) \in \mathbb{R}^{n \times d}$. $h_r$ is employed to project the rhythm embeddings $\Omega$ to the embedding space, followed by a $\tanh$ non-linearity such that $h_r(\Omega) \in \mathbb{R}^{n \times d}$. Gating matrix $\mathbf{W}$ is obtained by linearly projecting $\left[\Theta, \Omega\right] \in \mathbb{R}^{n \times 2d}$ using $h_c$ where $\left[\Theta, \Omega\right] \in \mathbb{R}^{n \times 2d}$ represents the concatenation of $\Theta \in \mathbf{R}^{n \times d}$ and $\Omega \in \mathbf{R}^{n \times d}$. Once $\left[\Theta, \Omega\right] \in \mathbb{R}^{n \times 2d}$ has been projected, a sigmoid function is employed to get the gating signal. Gated fusion can be represented as follows: 

\vspace{-12pt}
\begin{gather}
    \Theta' = \tanh(h_p(\Theta)) \label{1} \\
    \Omega' = \tanh(h_r(\Omega)) \label{2} \\
    \mathbf{W} = \sigma(h_c(\left[\Theta, \Omega \right])) \label{3}\\
    \mathbf{M} = \mathbf{W}\odot\Theta' + (1-\mathbf{W})\odot\Omega'
\end{gather}

where $\Theta'$ and $\Omega'$ represent the linear projections of $\Theta$ and $\Omega$. $\sigma$ represents the \textit{sigmoid} function, and $\mathbf{W}$ is the gating signal that is employed to fuse the updated position and user's interaction rhythm to get the fused matrix, $\mathbf{M} \in \mathbb{R}^{n \times d}$.
Figure~\ref{fig:interposgf} illustrates the INTERPOS-GF architecture.

\section{Experiments}
%In this section, we will discuss the datasets, baselines, and experimental setup. 

% \input{src/tables/tbl_dataset_stats}
\begin{table*}[t!]
\centering
\caption{Results on {\mbraction} dataset, best results are shown in bold, second best results are underlined.}
\vspace{-5pt}
%\resizebox{17cm}{!}{
\begin{tabular}
{l|cc|ccc|ccc}
\toprule
\multicolumn{1}{c|} {\multirow{2}{*}{\textbf{Category}}} & \multicolumn{2}{c|} {\multirow{2}{*}{\textbf{Method $\downarrow$ Metric $\xrightarrow{}$}}}                                                                    & \multicolumn{3}{|c|}{\textbf{NDCG}}                                                                   & \multicolumn{3}{c}{\textbf{HIT}}                                                                    \\ 
& &                                                                                 & \textbf{@10}                    & \textbf{@15}                    & \textbf{@20}                        & \textbf{@10}                    & \textbf{@15}                    & \textbf{@20}                    \\ \hline
\textbf{Popularity} & \multicolumn{2}{l|}{Pop}                                                                    & 0.0111                          & 0.0132                          & 0.0144                          &  0.0248                          & 0.0325                          & 0.0377                          \\ \hline
{\multirow{7}{*}{\textbf{Sequential}}} & \multicolumn{2}{l|}{GRU4Rec~\cite{tan2016improved}}                                                                & 0.0130                          & 0.0166                          & 0.0197                            & 0.0295                          & 0.0432                          & 0.0565                          \\
& \multicolumn{2}{l|}{LightSANs~\cite{fan2021lighter}}                                                              & 0.0142                          & 0.0181                          & 0.0217                         & 0.0312                          & 0.0458                          & 0.0609                          \\
& \multicolumn{2}{l|}{SASRec~\cite{kang2018self}}                                                                 & 0.0134                          & 0.0171                          & 0.0202                           & 0.0291                          & 0.0432                          & 0.0564                          \\
 & \multicolumn{2}{l|}{SINE~\cite{tan2021sparse}}                                                                   & 0.0091                          & 0.0119                          & 0.0142                            & 0.0202                          & 0.0307                          & 0.0403                          \\
& \multicolumn{2}{l|}{HGN~\cite{ma2019hierarchical}}                                                                    & 0.0128                          & 0.0165                          & 0.0199                          & 0.0283                          & 0.0425                          & 0.0569                          \\
 & \multicolumn{2}{l|}{GCSAN~\cite{xu2019graph}}                                                                  & 0.0129                          & 0.0164                          & 0.0195                          &  0.0281                          & 0.0412                          & 0.0546                          \\
\textbf{} & \multicolumn{2}{l|}{BERT4Rec~\cite{sun2019bert4rec}}                                                               & 0.0096                          & 0.0128                          & 0.0154                          & 0.0218                          & 0.0339                          & 0.0452                          \\\hline
{\multirow{2}{*}{\textbf{Time-aware}}} & \multicolumn{2}{l|}{FEARec~\cite{du2023frequency}} & 0.0147 & 0.0194 & 0.0201 & 0.0293 &0.0461 & 0.0514\\
\textbf{} & \multicolumn{2}{l|}{TiSASRec~\cite{li2020time}} & 0.0150 & 0.0199 & 0.0212 & 0.033 & 0.0514 & 0.0575  \\ \hline
\multicolumn{1}{c|}{\multirow{6}{*}{\textbf{This work}}} &
\multicolumn{1}{c}{\multirow{3}{*}{\textbf{LightSANs}}} & \multicolumn{1}{c|}{\textbf{INTERPOS-BF}} & \underline{0.0386} & \textbf{0.0465}                 & \textbf{0.0533}                 &  \textbf{0.0809}                 & \textbf{0.1107}                 & \textbf{0.1395}                 \\
& \multicolumn{1}{c}{}                                    & \multicolumn{1}{c|}{\textbf{INTERPOS-GF}} & 0.0385                          & 0.0461                          & 0.0523                          & 0.0790                          & 0.1077                          & 0.1340                          \\
& \multicolumn{1}{c}{}                                    & \multicolumn{1}{c|}{\textbf{INTERPOS-MF}} & \textbf{0.0387}                 & \underline{0.0463} & \underline{0.0527}          & \underline{0.0804} & \underline{0.1095} & \underline{0.1363} \\ \cline{2-9} 
& \multicolumn{1}{c}{\multirow{3}{*}{\textbf{SASRec}}}    & \multicolumn{1}{c|}{\textbf{INTERPOS-BF}} & 0.0234                          & 0.0286                          & 0.0331                          &  0.0491                          & 0.069                           & 0.0878                          \\
& \multicolumn{1}{c}{}                                    & \multicolumn{1}{c|}{\textbf{INTERPOS-GF}} & 0.0257                          & 0.0316                          & 0.0364                          &  0.0539                          & 0.0761                          & 0.0964                          \\
& \multicolumn{1}{c}{}                                    & \multicolumn{1}{c|}{\textbf{INTERPOS-MF}} & 0.0262                          & 0.0324                          & 0.0375                          & 0.0555                          & 0.0788                          & 0.1008    \\
\bottomrule
\end{tabular}
\label{tbl:action_results}
\vspace{-2pt}
\end{table*}

\begin{table*}[t!]
\centering
\caption{Results on {\mbrrole} dataset, best results are shown in bold, second best results are underlined.}
\vspace{-5pt}
%\resizebox{17cm}{!}{
\begin{tabular}{l|cc|ccc|ccc}
\toprule
\multicolumn{1}{c|} {\multirow{2}{*}{\textbf{Category}}} & \multicolumn{2}{c|} {\multirow{2}{*}{\textbf{Method $\downarrow$ Metric $\xrightarrow{}$}}}                                                                    & \multicolumn{3}{|c|}{\textbf{NDCG}}                                                                   & \multicolumn{3}{c}{\textbf{HIT}}                                                                    \\ 
& &                                                                                 & \textbf{@10}                    & \textbf{@15}                    & \textbf{@20}                        & \textbf{@10}                    & \textbf{@15}                    & \textbf{@20}                    \\ \hline
\textbf{Popularity} & \multicolumn{2}{l|}{Pop}                                                           & 0.0056                           & 0.0084                           & 0.0130                           & 0.0132                           & 0.0237                           & 0.0436               \\ \hline
{\multirow{7}{*}{\textbf{Sequential}}} & \multicolumn{2}{l|}{GRU4Rec~\cite{tan2016improved}}                               & 0.0174                           & 0.0210                           & 0.0246                          & 0.0362                           & 0.0497                           & 0.0651                             \\
& \multicolumn{2}{l|}{LightSANs~\cite{fan2021lighter}}                                                              & 0.0165                           & 0.0206                           & 0.0238                          & 0.0357                           & 0.0511                           & 0.0649                                                 \\
& \multicolumn{2}{l|}{SASRec~\cite{kang2018self}}                                                                 & 0.0146                           & 0.018                            & 0.0212                           & 0.0522                           & 0.0717                           & 0.0916                                               \\
& \multicolumn{2}{l|}{SINE~\cite{tan2021sparse}}                                                                   & 0.0094                           & 0.0112                           & 0.0131                           & 0.0192                           & 0.0262                           & 0.0343                                                \\
& \multicolumn{2}{l|}{HGN~\cite{ma2019hierarchical}}                                                                    & 0.0126                           & 0.0160                           & 0.0190                           & 0.0270                           & 0.0399                           & 0.0523                                              \\
& \multicolumn{2}{l|}{GCSAN~\cite{xu2019graph}}                                                                  & 0.0120                           & 0.0149                           & 0.0175                           & 0.0258                           & 0.0365                           & 0.0477                                                \\
& \multicolumn{2}{l|}{BERT4Rec~\cite{sun2019bert4rec}}                                                               & 0.0102                           & 0.0128                           & 0.0151                           & 0.0210                           & 0.0308                           & 0.0403                                               \\\hline
{\multirow{2}{*}{\textbf{Time-aware}}} & \multicolumn{2}{l|}{FEARec~\cite{du2023frequency}} & 0.0181 & 0.0191 & 0.024 & 0.04 & 0.0558 & 0.0651\\
\textbf{} & \multicolumn{2}{l|}{TiSASRec~\cite{li2020time}} & 0.0194 & 0.0206 & 0.0203 & 0.0387 & 0.0522 & 0.0654  \\ \hline
\multicolumn{1}{c|}{\multirow{6}{*}{\textbf{This work}}} &
\multicolumn{1}{c|}{\multirow{3}{*}{\textbf{LightSANs}}} & \multicolumn{1}{c|}{\textbf{INTERPOS-BF}} & 0.0328                           & 0.0398                           & 0.0451                            & 0.0684                           & 0.0949                           & 0.1174                            \\
& \multicolumn{1}{c|}{}                                    & \multicolumn{1}{c|}{\textbf{INTERPOS-GF}} & \textbf{0.0353 }                          & \textbf{0.0429 }                          & \textbf{0.0485 }                          & \textbf{0.0747}                           & \textbf{0.1035}                          & \textbf{0.1271}                                                 \\
& \multicolumn{1}{c|}{}                                    & \multicolumn{1}{c|}{\textbf{INTERPOS-MF}} & \underline{0.0341}                           & \underline{0.0412}                          & \underline{0.0466}                       & \underline{0.0724}                           & \underline{0.0996}                           & \underline{0.1220}                                   \\ \cline{2-9}
& \multicolumn{1}{c|}{\multirow{3}{*}{\textbf{SASRec}}}    & \multicolumn{1}{c|}{\textbf{INTERPOS-BF}} & 0.0252                           & 0.0303                           & 0.0350                           & 0.0522                           & 0.0717                           & 0.0916                                   \\
& \multicolumn{1}{c|}{}                                    & \multicolumn{1}{c|}{\textbf{INTERPOS-GF}} & 0.0294                           & 0.0354                           & 0.0402                            & 0.0612                           & 0.084                            & 0.1042                                              \\
& \multicolumn{1}{c|}{}                                    & \multicolumn{1}{c|}{\textbf{INTERPOS-MF}} & 0.0283                           & 0.0342                           & 0.0387                           & 0.0597                           & 0.0821                           & 0.1012                                              \\
\bottomrule
\end{tabular}
\label{tbl:role}
\vspace{-12pt}
\end{table*}

\subsection{Experimental Setup}
We use RecBole~\cite{recbole[1.0], recbole[1.2.0], recbole[2.0]} for the implementation.
We integrate {\ourmodel} into LightSANs~\cite{fan2021lighter} and SASRec~\cite{kang2018self} architectures.
%use RecBole~\cite{recbole[1.0], recbole[1.2.0], recbole[2.0]} for the implementation. 
% In this work, we integrate INTERPOS into \textit{LightSANs}~\cite{fan2021lighter}. 
LightSANs integration is as follows.
We use two transformer layers with two attention heads, the hidden size is 64 and the inner size is 256, while the number of latent interests is set to 5.
% The maximum sequence length is 50. 
% We have a hidden dropout probability of 0.5 and an attention dropout probability of 0.5. 
The sequence length is capped at 50, with a hidden dropout probability of 0.5 and an attention dropout probability of 0.5.
Hidden activation is Gaussian Error Linear Unit (GELU) and cross-entropy (CE) loss for the next-item prediction task. 
Models are trained for 100 epochs with an early-stopping patience of 10 with a batch size of 4096, and our learning rate is 0.001. 
SASRec has two transformer layers with two attention heads; the hidden size is 128 and the inner size is 256. We employ GELU activation and the \textit{Adam} optimizer with a learning rate of 0.001. 
The same base SASRec architecture is consistently kept across for {\ourmodel} integration to get INTERPOS-BM, INTERPOS-GF, and INTERPOS-MF.
The models' trainable parameters depend on the maximum interaction rhythm difference. 
A leave-one-out strategy is employed for validation and testing.
The full item set is used for evaluation. 
We normalize the interaction rhythms in {\mbraction}, {\mbrcasual}, {\mbrrole}, {\mbrpuzzle}, {\mbrsimulation}, {\mbrstrategy} with 0.2 and in case of {\mbr} dataset, we clip the interaction rhythm at 800. 
We employ \textit{HIT@K} and \textit{NDCG@K} evaluation metrics where $k \in \left\{10, 15, 20\right\}$. %SASRec has two transformer layers with two attention heads; the hidden size is 128 and the inner size is 256. We employ GELU activation and the \textit{Adam} optimizer with a learning rate of 0.001. The same base SASRec architecture is consistently kept across for {\ourmodel} integration to get INTERPOS-BM, INTERPOS-GF, and INTERPOS-MF.

%We employ cuda 11.5 for our experiments on {\mbraction}, {\mbrcasual}, {\mbrrole}, {\mbrpuzzle}, {\mbrsimulation} and {\mbrstrategy}. 

\subsection{Datasets}
Our experiments use 7 datasets for evaluating {\ourmodel} and benchmarking the performance gains in comparison with the competing baselines. 
MobileRec~\cite{maqbool2023mobilerec} is a large-scale dataset with over 19 million user-app interactions spanning 48 categories. 
% Concerning the number of interactions, 
We select the top 6 categories with the highest number of interactions, including ({\mbraction}, {\mbrrole}, {\mbrpuzzle}, {\mbrcasual}, {\mbrsimulation}, {\mbrstrategy}, and {\mbrsimulation}).
% have been selected for conducting the experiments in this work. 
On top of these 6 datasets, a full-scale MobileRec~\cite{maqbool2023mobilerec} dataset with all 48 categories has also been used to establish the efficacy of {\ourmodel}. 
%There are 80k unique users, 529 unique applications, around 9 average interactions per user, 1502 average interactions per item, and 0.79M user/item interactions in the {\mbraction} dataset. 
% Table \ref{tbl_dataset_stats} can be consulted for viewing the datasets' statistics. 
Table \ref{tbl_dataset_stats} presents detailed statistics of datasets.

\subsection{Baselines}
We have employed several strong baselines for benchmarking the efficacy of {\ourmodel}, as described in the following. 

{Pop} is a simple popularity-based recommender system. This model captures the popularity of items in the dataset and suggests the most popular items to users as recommendations. \textbf{GRU4Rec}~\cite{tan2016improved} applies an RNN-based method for the session-based recommendation.
This baseline presents a method to account for data distribution shifts along with data augmentation. \textbf{LightSANs}~\cite{fan2021lighter}
 introduces a self-attention network with low-rank decomposition that projects users' historical items onto a small number of latent interests. \textbf{SASRec}~\cite{kang2018self} employs the attention mechanism for sequential recommendation task. \textbf{SINE}~\cite{tan2021sparse} proposes to use multiple embeddings to capture various aspects of a user’s behavior. \textbf{HGN}~\cite{ma2019hierarchical} emphasizes the importance of recent chronological user-app interactions and integrates Bayesian Personalized Ranking (BPR) to capture both long-term and short-term user interests. \textbf{GCSAN}~\cite{xu2019graph} a graph-contextualized self-attention model is proposed that employs both graph neural networks and self-attention networks. 
Graph neural network captures rich local associations, while self-attention networks capture long-range correlations. \textbf{BERT4Rec}~\cite{sun2019bert4rec} utilizes bidirectional self-attention, framing the sequential recommendation problem under the cloze objective. \textbf{TiSASRec}~\cite{li2020time} incorporates time intervals between user interactions by explicitly modeling the timestamp of the interactions in the self-attention layer. \textbf{FEARec}~\cite{du2023frequency} explicitly learns low-frequency and high-frequency information and combines time and frequency characteristics via auto-correlation.

\section{Results and Discussion}
%In this section, we will discuss the quantitative and qualitative results and also do a comparative analysis of {\ourmodel} concerning various baselines.
% \subsection{Quantitative Results}

We observe that {\ourmodel} with all the variations, exhibits a strong capability to steer the target model towards better performance. 
Our experiments also show that the proposed rhythm integration strategies INTERPOS-BF, INTERPOS-GF, and INTERPOS-MF show their strengths on all datasets.

%Table~\ref{tbl_dataset_stats} depicts the distributional variations among different categorical splits of {\mbr} dataset. 
%For example, {\mbraction} dataset has $\approx$81k unique users and 529 unique apps with nearly 1500 interactions per unique app and around 9 interactions per unique user on average. 
\stitle{Performance on Action dataset.}
{\mbraction} dataset has $\approx$81k unique users and 529 unique apps with nearly 1500 interactions per unique app and around 9 interactions per unique user on average. 
We report our results for {\mbraction} dataset in Table~\ref{tbl:action_results}.
Overall, when {\ourmodel} is incorporated in the LightSANs, it performs better than SASRec integration.  
In LightSANs integration, INTERPOS-BF performs better on \textit{NDCG@k} where $k \in \left\{15, 20\right\}$. 
On \textit{HIT@k}, INTERPOS-BF is a better-performing strategy for $k \in \left\{10, 15, 20\right\}$. 
% while INTERPOS-MF outclasses both INTERPOS-BF and INTERPOS-GF on \textit{NDCG@10}. 
This observation that within one dataset, our proposed fusion strategies find their strengths on different values of $k$ in both \textit{NDCG} and \textit{HIT} can be attributed to relative metric strictness. 
% For instance, INTERPOS-GF emerges to be a better choice for \textit{NDCG@5} on {\mbraction} dataset. 
We argue that the choice of INTERPOS (i.e.,  INTERPOS-BF, INTERPOS-GF, INTERPOS-MF) fusion strategies depends upon the dataset's distributional variations and the underlying metric. 
In comparison with other competing baselines, we observe that INTERPOS outperforms all the baselines by a significant margin. 
Among the baselines, time-aware baselines perform the best while the popularity remains the worst.
The sequential recommendation baselines have mixed results among all the sequential baselines. \textit{LightSANs}~\cite{fan2021lighter} demonstrates strong performance on \textit{NDCG@k} and \textit{HIT@k} for all values of \textit{k}.
The time-aware baselines outperform all the sequential baselines except for the \textit{NDCG@20} metric, which shows that time-sensitivity in fact plays an important role. 
INTERPOS manages to outperform time-aware baselines by a large margin. 
As compared to best-performing baselines, on \textit{NDCG@10}, we observe a percentage improvement of 157.3\%, 156.67\%, and 158\% by INTERPOS-BF, INTERPOS-GF, and INTERPOS-MF respectively in \textit{LightSANs} integration. 
On \textit{NDCG@20}, INTERPOS-BF, INTERPOS-GF, and INTERPOS-MF manage to achieve 145.62\%, 141.01\%, and 142.85\% improvement compared to \textit{LightSANs}~\cite{fan2021lighter}, respectively. 
On \textit{HIT@10} and \textit{HIT@20}, INTERPOS-BF manages a percentage improvement of 145.15\% and 142.6\%, INTERPOS-GF demonstrates an improvement of 139.39\% and 133.04\% while INTERPOS-MF shows 143.63\% and 137.04\% improvement. 
% INTERPOS demonstrate improvements for all values of $k$ on \textit{NDCG@k} but we will only focus on $k \in \left\{5,10 \right\}$. 

\begin{table*}[t!]
\centering
\caption{Results on {\mbrcasual} dataset, best results are shown in bold, second best results are underlined.}
\vspace{-4pt}
%\resizebox{17cm}{!}{
\begin{tabular}
{l|cc|ccc|ccc}
\toprule
\multicolumn{1}{c|} {\multirow{2}{*}{\textbf{Category}}} & \multicolumn{2}{c|} {\multirow{2}{*}{\textbf{Method $\downarrow$ Metric $\xrightarrow{}$}}}                                                                    & \multicolumn{3}{|c|}{\textbf{NDCG}}                                                                   & \multicolumn{3}{c}{\textbf{HIT}}                                                                    \\ 
& &                                                                                 & \textbf{@10}                    & \textbf{@15}                    & \textbf{@20}                        & \textbf{@10}                    & \textbf{@15}                    & \textbf{@20}                    \\ \hline
\textbf{Popularity} & \multicolumn{2}{l|}{Pop} & 0.0112                           & 0.0144                           & 0.0186                          & 0.0204                           & 0.0324                           & 0.0501                          \\ \hline
{\multirow{7}{*}{\textbf{Sequential}}} & \multicolumn{2}{l|}{GRU4Rec~\cite{tan2016improved}}                                                                & 0.0294                           & 0.0351                           & 0.0399           & 0.0576                           & 0.0791                           & 0.0998                                               \\
& \multicolumn{2}{l|}{LightSANs~\cite{fan2021lighter}}                                                              & 0.0294                           & 0.0355                           & 0.0403                           &  0.0559                           & 0.0793                           & 0.0996                                                \\
& \multicolumn{2}{l|}{SASRec~\cite{kang2018self}}                                                                 & 0.0283                           & 0.0340                           & 0.0386                           &   0.0566                           & 0.0783                           & 0.0975                      \\
& \multicolumn{2}{l|}{SINE~\cite{tan2021sparse}}                                                                   & 0.0256                           & 0.0295                           & 0.0327                           &  0.0466                           & 0.0613                           & 0.0752                              \\
& \multicolumn{2}{l|}{HGN~\cite{ma2019hierarchical}}                                                                    & 0.0288                           & 0.0343                           & 0.0392                                 & 0.0569                           & 0.0778                           & 0.0985                           \\
& \multicolumn{2}{l|}{GCSAN~\cite{xu2019graph}}                                                                  & 0.0267                           & 0.0317                           & 0.0361                           & 0.0508                           & 0.0699                           & 0.0885                      \\
& \multicolumn{2}{l|}{BERT4Rec~\cite{sun2019bert4rec}}                                                               & 0.0252                           & 0.0295                           & 0.0331                           &  0.0515                           & 0.0678                           & 0.0830                       \\\hline
{\multirow{2}{*}{\textbf{Time-aware}}} & \multicolumn{2}{l|}{FEARec~\cite{du2023frequency}} & 0.0304 & 0.041 & 0.0353 & 0.0599 & 0.089 & 0.0973\\
\textbf{} & \multicolumn{2}{l|}{TiSASRec~\cite{li2020time}} & 0.0261 & 0.04 & 0.0344 & 0.057 & 0.0822 & 0.1023  \\ \hline
\multicolumn{1}{c|}{\multirow{6}{*}{\textbf{This work}}} &
\multicolumn{1}{c|}{\multirow{3}{*}{\textbf{LightSANs}}} & \multicolumn{1}{c|}{\textbf{INTERPOS-BF}} & 0.0390                           & 0.0474                           & \underline{0.0551}                         & 0.0790                           & 0.1108                           & \underline{0.1434}                           \\
& \multicolumn{1}{c|}{}                                    & \multicolumn{1}{c|}{\textbf{INTERPOS-GF}} & \underline{0.0400}                           & \underline{0.0484}                          & 0.0545                            & \underline{0.0817}                           & \underline{0.1136}                        & 0.1394                           \\
& \multicolumn{1}{c|}{}                                    & \multicolumn{1}{c|}{\textbf{INTERPOS-MF}} & \textbf{0.0424}                           & \textbf{0.0518 }                          & \textbf{0.0592 }                         & \textbf{0.0871}                           & \textbf{0.123}                            & \textbf{0.1541}                              \\ \cline{2-9}
& \multicolumn{1}{c|}{\multirow{3}{*}{\textbf{SASRec}}}    & \multicolumn{1}{c|}{\textbf{INTERPOS-BF}} & 0.0316                           & 0.0377                           & 0.0435                          & 0.0634                           & 0.0867                           & 0.1113                             \\
& \multicolumn{1}{c|}{}                                    & \multicolumn{1}{c|}{\textbf{INTERPOS-GF}} & 0.0343                           & 0.0402                           & 0.0455                          & 0.0664                           & 0.0891                           & 0.1112                          \\
& \multicolumn{1}{c|}{}                                    & \multicolumn{1}{c|}{\textbf{INTERPOS-MF}} & 0.0329                           & 0.0392                           & 0.0445                           &  0.0654                           & 0.0894                           & 0.1119                            \\ \bottomrule  
\end{tabular}

\label{tbl:casual}
%\vspace{-2pt}
\end{table*}
\begin{table*}[t!]
\centering
\caption{Results on {\mbrstrategy} dataset, best results are shown in bold, second best results are underlined.}
\vspace{-4pt}
%\resizebox{17cm}{!}{
\begin{tabular}
{l|cc|ccc|ccc}
\toprule
\multicolumn{1}{c|} {\multirow{2}{*}{\textbf{Category}}} & \multicolumn{2}{c|} {\multirow{2}{*}{\textbf{Method $\downarrow$ Metric $\xrightarrow{}$}}}                                                                    & \multicolumn{3}{|c|}{\textbf{NDCG}}                                                                   & \multicolumn{3}{c}{\textbf{HIT}}                                                                    \\ 
& &                                                                                 & \textbf{@10}                    & \textbf{@15}                    & \textbf{@20}                        & \textbf{@10}                    & \textbf{@15}                    & \textbf{@20}                    \\ \hline
\textbf{Popularity} & \multicolumn{2}{l|}{Pop} & 0.0194                           & 0.0213                           & 0.0244                           & 0.0341                           & 0.0411                           & 0.0543 \\ \hline
{\multirow{7}{*}{\textbf{Sequential}}} & \multicolumn{2}{l|}{GRU4Rec~\cite{tan2016improved}}                                                                         & 0.0212                           & 0.0270                           & 0.0313                           &0.0478                           & 0.0698                           & 0.0880                   \\
& \multicolumn{2}{l|}{LightSANs~\cite{fan2021lighter}}                                                                       & 0.0268                           & 0.0332                           & 0.0384                           &0.0579                           & 0.082                            & 0.1040                    \\
& \multicolumn{2}{l|}{SASRec~\cite{kang2018self}}                                                                          & 0.0186                           & 0.0240                           & 0.0285                           & 0.0412                           & 0.0618                           & 0.0808 \\
& \multicolumn{2}{l|}{SINE~\cite{tan2021sparse}}                                                                            & 0.0098                           & 0.0127                           & 0.0156                           &0.0200                           & 0.0308                           & 0.0432 \\
& \multicolumn{2}{l|}{HGN~\cite{ma2019hierarchical}}                                                                             & 0.0207                           & 0.0265                           & 0.0312                           &0.0445                           & 0.0664                           & 0.0863\\
& \multicolumn{2}{l|}{GCSAN~\cite{xu2019graph}}                                                                           & 0.0171                           & 0.0216                           & 0.0255                           &0.0387                           & 0.0559                           & 0.0722\\
& \multicolumn{2}{l|}{BERT4Rec~\cite{sun2019bert4rec}}                                                                        & 0.0140                           & 0.0178                           & 0.0208                          & 0.0307                           & 0.0454                           & 0.0580   \\ \hline
{\multirow{2}{*}{\textbf{Time-aware}}} & \multicolumn{2}{l|}{FEARec~\cite{du2023frequency}} & 0.0237 & 0.0336 & 0.0336 & 0.0599 & 0.0899 & 0.0943\\
\textbf{} & \multicolumn{2}{l|}{TiSASRec~\cite{li2020time}} & 0.0253 & 0.0375 & 0.0375 & 0.0592 & 0.0912 & 0.1056  \\ \hline
\multicolumn{1}{c|}{\multirow{6}{*}{\textbf{This work}}} &
\multicolumn{1}{c|}{\multirow{3}{*}{\textbf{LightSANs}}} & \multicolumn{1}{c|}{\textbf{INTERPOS-BF}} & \textbf{0.0403}                           & \textbf{0.0482}                           & \textbf{0.0552}                            &\textbf{ 0.086}                            & \textbf{0.116}                            & \textbf{0.1457}                           \\
& \multicolumn{1}{c|}{}                                    & \multicolumn{1}{c|}{\textbf{INTERPOS-GF}} & \underline{0.0395}                           & \underline{0.0471}                          & \underline{0.0540}                        & \underline{0.0845}                           & \underline{0.1133}                        & \underline{0.1423}   \\
& \multicolumn{1}{c|}{}                                    & \multicolumn{1}{c|}{\textbf{INTERPOS-MF}} & 0.0384                           & 0.0465                           & 0.0533                           &  0.0816                           & 0.1122                           & 0.1411  \\ \cline{2-9}
& \multicolumn{1}{c|}{\multirow{3}{*}{\textbf{SASRec}}}    & \multicolumn{1}{c|}{\textbf{INTERPOS-BF}} & 0.0283                           & 0.0348                           & 0.0404                            & 0.0611                           & 0.0857                           & 0.1093      \\
& \multicolumn{1}{c|}{}                                    & \multicolumn{1}{c|}{\textbf{INTERPOS-GF}} & 0.0299                           & 0.0368                           & 0.0426                           & 0.064                            & 0.0902                           & 0.1150  \\
& \multicolumn{1}{c|}{}                                    & \multicolumn{1}{c|}{\textbf{INTERPOS-MF}} & 0.0316                           & 0.0383                           & 0.0442                           &0.0661                           & 0.0914                           & 0.1165  \\
\bottomrule
\end{tabular}
\label{tbl:strategy}
\vspace{-6pt}
\end{table*}

\stitle{Performance on RolePlaying dataset.}
{\mbrrole} has roughly 78k unique users and 658 unique apps, with an average interaction of 1139.57 per unique app and 9.63 per unique user. 
Table~\ref{tbl:role} reports the results for {\mbrrole}. 
INTERPOS-MF is the best-performing fusion strategy among INTERPOS when integrated into \textit{LightSANs}. 
Again, time-aware baselines perform better as compared to other baselines.
Among sequential baselines, \textit{GRU4Rec}~\cite{tan2016improved} is the among all the baselines. 
INTERPOS-MF integration with \textit{LightSANs} obtains a considerable percentage improvement over all baselines. 
INTERPOS-GF integration with \textit{LightSANs} achieves a significant 88.62\% improvement against \textit{GRU4Rec}~\cite{tan2016improved} on \textit{NDCG@10}.
Similarly, it demonstrates an 81.95\% improvement on \textit{NDCG@10} compared to \textit{TiSASRec}. 
On \textit{HIT@15} and \textit{HIT@20}, INTERPOS-GF shows 85.48\% and 94.34\% improvement in comparison with time-aware baselines.

% Casual Dataset
\stitle{Performance on Casual dataset.}
Table~\ref{tbl:casual} summarizes the results on {\mbrcasual} dataset. 
INTERPOS-MF integrated with \textit{LightSANs} outclasses baselines and other fusion strategies. 
Time-aware baselines perform better as compared to others with the exception of \textit{NDCG@20} metric, where \textit{LightSANs}~\cite{fan2021lighter} performs better as compared to other baselines.
% on \textit{NDCG@5} metric, \textit{GRU4Rec}~\cite{tan2016improved} and \textit{LightSANs}~\cite{fan2021lighter} perform equally well. 
% INTERPOS-MF outclasses both of these competing baselines on \textit{NDCG@10} by demonstrating an improvement of 39.47\%. 
% on \textit{NDCG@10} evaluation metric, we observe that both baseline methods show equally good performance. 
INTREPOS-MF integrated with \textit{LightSANs} manages to outperform the best time-aware baseline (i.e., \textit{FEARec}) by 39.47\% on \textit{NDCG@10} metric and by 67.70\% on \textit{NDCG@20}. 
Notable improvements are obtained by other fusion strategies on \textit{NDCG@10} and \textit{NDCG@15}. 
Consider the \textit{NDCG@10} and \textit{NDCG@15} metrics, INTERPOS-BF manages to outperform \textit{FEARec} by 28.28\% and 15.60\% while INTERPOS-GF demonstrates a percent improvement of 31.57\% and 18.04\%. 
On \textit{HIT@10}, \textit{HIT@15}, and \textit{HIT@20}, INTERPOS-MF shows 45.40\%, 38.20\%, and 50.63\% improvements against the best-performing (i.e., time-aware) baselines on these metrics. 
INTERPOS-BF shows an improvement of 31.88\%, 24.49\%, and 40.17\% on these metrics in comparison with the best-performing baselines. 
Similarly, INTERPOS-GF manages to obtain, 36.39\%, 27.64\%, and 36.26\% improvement in comparison with the best performing baselines on \textit{HIT@10}, and \textit{HIT@15}, and \textit{HIT@20}.

\begin{table*}[t!]
\centering
\caption{Results on {\mbrsimulation} dataset, best results are shown in bold, second best results are underlined.}
\vspace{-4pt}
%\resizebox{17cm}{!}{

\begin{tabular}
{l|cc|ccc|ccc}
\toprule
\multicolumn{1}{c|} {\multirow{2}{*}{\textbf{Category}}} & \multicolumn{2}{c|} {\multirow{2}{*}{\textbf{Method $\downarrow$ Metric $\xrightarrow{}$}}}                                                                    & \multicolumn{3}{|c|}{\textbf{NDCG}}                                                                   & \multicolumn{3}{c}{\textbf{HIT}}                                                                    \\ 
& &                                                                                 & \textbf{@10}                    & \textbf{@15}                    & \textbf{@20}                        & \textbf{@10}                    & \textbf{@15}                    & \textbf{@20}                    \\ \hline
\textbf{Popularity} & \multicolumn{2}{l|}{Pop}  & 0.0136                           & 0.0178                           & 0.0202                            & 0.0255                           & 0.0414                           & 0.0518                          \\ \hline
{\multirow{7}{*}{\textbf{Sequential}}} & \multicolumn{2}{l|}{GRU4Rec~\cite{tan2016improved}}                                                                         & 0.0197                           & 0.0245                           & 0.0289                              & 0.0407                           & 0.0587                           & 0.0774                             \\
& \multicolumn{2}{l|}{LightSANs~\cite{fan2021lighter}}                                                                       & 0.0198                           & 0.0248                           & 0.0292                            & 0.0412                           & 0.0601                           & 0.0789                           \\
& \multicolumn{2}{l|}{SASRec~\cite{kang2018self}}                                                                          & 0.0175                           & 0.0221                           & 0.0263                          & 0.0383                           & 0.0558                           & 0.0732                           \\
& \multicolumn{2}{l|}{SINE~\cite{tan2021sparse}}                                                                            & 0.0145                           & 0.0171                           & 0.0200                          & 0.0345                           & 0.0445                           & 0.0567                          \\
& \multicolumn{2}{l|}{HGN~\cite{ma2019hierarchical}}                                                                             & 0.0181                           & 0.0231                           & 0.0273                           & 0.0383                           & 0.0571                           & 0.0751                            \\
& \multicolumn{2}{l|}{GCSAN~\cite{xu2019graph}}                                                                           & 0.0179                           & 0.0226                           & 0.0268                           & 0.0397                           & 0.0579                           & 0.0755                            \\
& \multicolumn{2}{l|}{BERT4Rec~\cite{sun2019bert4rec}}                                                                        & 0.0149                           & 0.0194                           & 0.0227                           &  0.0324                           & 0.0495                           & 0.0633                                   \\\hline
{\multirow{2}{*}{\textbf{Time-aware}}} & \multicolumn{2}{l|}{FEARec~\cite{du2023frequency}} & 0.0245 & 0.0246 & 0.025 & 0.0417 & 0.0644 & 0.0731\\
\textbf{} & \multicolumn{2}{l|}{TiSASRec~\cite{li2020time}} & 0.0159 & 0.0313 & 0.0285 & 0.0455 & 0.0611 & 0.0781  \\ \hline
\multicolumn{1}{c|}{\multirow{6}{*}{\textbf{This work}}} & \multicolumn{1}{c|}{\multirow{3}{*}{\textbf{LightSANs}}} & \multicolumn{1}{c|}{\textbf{INTERPOS-BF}} & \textbf{0.036}                            & \textbf{0.0435}                           & \textbf{0.0505}                          & \textbf{0.0694}                           & \textbf{0.0977}                           & \textbf{0.1277}                           \\
& \multicolumn{1}{c|}{}                                    & \multicolumn{1}{c|}{\textbf{INTERPOS-GF}} & 0.0299                           & 0.0369                           & 0.0429                          & 0.0593                           & 0.0857                           & 0.1113 \\
& \multicolumn{1}{c|}{}                                    & \multicolumn{1}{c|}{\textbf{INTERPOS-MF}} & \underline{0.0335}                           & \underline{0.0409}                        & \underline{0.0475}                            & \underline{0.0674}                        & \underline{0.0953}                           & \underline{0.1232}                         \\ \cline{2-9}
& \multicolumn{1}{c|}{\multirow{3}{*}{\textbf{SASRec}}}    & \multicolumn{1}{c|}{\textbf{INTERPOS-BF}} & 0.0177                           & 0.0223                           & 0.0261                           &0.0389                           & 0.0564                           & 0.0724   \\
& \multicolumn{1}{c|}{}                                    & \multicolumn{1}{c|}{\textbf{INTERPOS-GF}} & 0.0216                           & 0.0274                           & 0.0327                          & 0.0462                           & 0.0680                           & 0.0905                           \\
& \multicolumn{1}{c|}{}                                    & \multicolumn{1}{c|}{\textbf{INTERPOS-MF}} & 0.0211                           & 0.0266                           & 0.0316                           & 0.0446                           & 0.0657                           & 0.0867\\
\bottomrule
\end{tabular}
\label{tbl:simulation}
\vspace{-2pt}
\end{table*}

\begin{table*}[t!]
\centering
\caption{Results on {\mbrpuzzle} dataset, best results are shown in bold, second best results are underlined.}
\vspace{-4pt}
%\resizebox{17cm}{!}{
\begin{tabular}
{l|cc|ccc|ccc}
\toprule
\multicolumn{1}{c|} {\multirow{2}{*}{\textbf{Category}}} & \multicolumn{2}{c|} {\multirow{2}{*}{\textbf{Method $\downarrow$ Metric $\xrightarrow{}$}}}                                                                    & \multicolumn{3}{|c|}{\textbf{NDCG}}                                                                   & \multicolumn{3}{c}{\textbf{HIT}}                                                                    \\ 
& &                                                                                 & \textbf{@10}                    & \textbf{@15}                    & \textbf{@20}                        & \textbf{@10}                    & \textbf{@15}                    & \textbf{@20}                    \\ \hline
\textbf{Popularity} & \multicolumn{2}{l|}{Pop} & 0.0172                           & 0.0215                           & 0.0244                    & 0.0382                           & 0.0545                           & 0.0668                         \\ \hline
{\multirow{7}{*}{\textbf{Sequential}}} & \multicolumn{2}{l|}{GRU4Rec~\cite{tan2016improved}}                                                                         & 0.0224                           & 0.0271                           & 0.0317                           & 0.0458                           & 0.0635                           & 0.0832                   \\
& \multicolumn{2}{l|}{LightSANs~\cite{fan2021lighter}}                                                                       & 0.0229                           & 0.0278                           & 0.0322                          & 0.0464                           & 0.0652                           & 0.0837  \\
& \multicolumn{2}{l|}{SASRec~\cite{kang2018self}}                                                                          & 0.0221                           & 0.0270                           & 0.0313                           & 0.0446                           & 0.0635                           & 0.0816   \\
& \multicolumn{2}{l|}{SINE~\cite{tan2021sparse}}                                                                            & 0.0194                           & 0.0227                           & 0.0257                           & 0.0377                           & 0.0506                           & 0.0629 \\
& \multicolumn{2}{l|}{HGN~\cite{ma2019hierarchical}}                                                                             & 0.0222                           & 0.0270                           & 0.0314                           &0.0454                           & 0.0638                           & 0.0823    \\
& \multicolumn{2}{l|}{GCSAN~\cite{xu2019graph}}                                                                           & 0.0227                           & 0.0274                           & 0.0313                           & 0.0465                           & 0.0644                           & 0.0812 \\
& \multicolumn{2}{l|}{BERT4Rec~\cite{sun2019bert4rec}}                                                                        & 0.0189                           & 0.0228                           & 0.0259                           & 0.0363                           & 0.0510                           & 0.0642  \\ \hline
{\multirow{2}{*}{\textbf{Time-aware}}} & \multicolumn{2}{l|}{FEARec~\cite{du2023frequency}} & 0.0267 & 0.0288 & 0.0338 & 0.052 & 0.0679 & 0.0725\\
\textbf{} & \multicolumn{2}{l|}{TiSASRec~\cite{li2020time}} & 0.0262 & 0.0316 & 0.0355 & 0.048 & 0.0717 & 0.0828  \\ \hline
\multicolumn{1}{c|}{\multirow{6}{*}{\textbf{This work}}} 
& \multicolumn{1}{c|}{\multirow{3}{*}{\textbf{LightSANs}}} & \multicolumn{1}{c|}{\textbf{INTERPOS-BF}} & \underline{0.0377}                           & \underline{0.0444}                          & 0.0505                           & \underline{0.0719}                           & \underline{0.0976}                           & 0.1234 \\
& \multicolumn{1}{c|}{}                                    & \multicolumn{1}{c|}{\textbf{INTERPOS-GF}} & \textbf{0.0408}                           & \textbf{0.0485}                           & \textbf{0.0556}                          & \textbf{0.0789}                           & \textbf{0.1082}                           & \textbf{0.1379}  \\
& \multicolumn{1}{c|}{}                                    & \multicolumn{1}{c|}{\textbf{INTERPOS-MF}} & 0.0370                           & 0.0440                           & \underline{0.0506}                          & 0.0701                           & 0.0967                           & \underline{0.1248}     \\ \cline{2-9}
& \multicolumn{1}{c|}{\multirow{3}{*}{\textbf{SASRec}}}    & \multicolumn{1}{c|}{\textbf{INTERPOS-BF}} & 0.0231                           & 0.0285                           & 0.0330                           &0.0478                           & 0.0683                           & 0.0872  \\
& \multicolumn{1}{c|}{}                                    & \multicolumn{1}{c|}{\textbf{INTERPOS-GF}} & 0.0283                           & 0.0344                           & 0.0401                          & 0.0583                           & 0.0814                           & 0.1055   \\
& \multicolumn{1}{c|}{}                                    & \multicolumn{1}{c|}{\textbf{INTERPOS-MF}} & 0.0246                           & 0.0303                           & 0.0354                           & 0.0515                           & 0.0731                           & 0.0949  \\
\bottomrule
\end{tabular}
\label{tbl:puzzle}
\vspace{-6pt}
\end{table*}

%Simulation Dataset
\stitle{Performance on Simulation dataset.}
Table~\ref{tbl:simulation} presents the results on {\mbrsimulation} dataset. 
On \textit{NDCG@20}, under LightSANs integration, INTERPOS-BF demonstrates 72.94\% improvement over the best-performing baseline  \textit{LightSANs}. 
INTERPOS-GF and INTERPOS-MF show 46.91\% and 62.67\% improvement over \textit{LightSANs} on \textit{NDCG@20}. 
On \textit{NDCG@10} \textit{FEARec} outclasses other baselines. 
INTERPOS-BF manages to get 46.93\% improvement over the best-performing baseline. 
On \textit{HIT@10} and \textit{HIT@20}, INTERPOS-BF outperforms the best baseline method \textit{TiSASRec} by 52.52\% and 62.67\%, while INTERPOS-GF manages a 30.32\% and 41.78\% improvement and INTERPOS-MF shows a 48.13\% and 57.74\% improvement, respectively.

%Strategy Dataset
\stitle{Performance on Strategy dataset.}
On {\mbrstrategy} dataset, INTERPOS in both integration with \textit{LightSANs} and \textit{SASRec} outclass all the competing baselines. 
Detailed results are shown in Table~\ref{tbl:strategy}. 
On \textit{NDCG@10}, \textit{LightSANs}~\cite{fan2021lighter} outclasses the other baseline methods. 
INTERPOS-BF under \textit{LightSANs} integration manages to outperform \textit{LightSANs}~\cite{fan2021lighter} on \textit{NDCG@10} by 50.37\%. 
Similarly, on \textit{HIT@15} and \textit{HIT@20}, INTERPOS-BF manages to outperform the best competing baselines \textit{FEARec} and \textit{LightSANs}~\cite{fan2021lighter} by 28.53\% and 43.75\%. 
On the other hand, on \textit{NDCG@10}, INTERPOS-GF and INTERPOS-MF show a percentage improvement of 47.38\% and 43.28\%, respectively. 
On \textit{NDCG@20}, INTERPOS-GF and INTERPOS-MF manage to get 40.62\% and 38.8\% improvement over \textit{LightSANs}~\cite{fan2021lighter}.
On \textit{HIT@10}, we notice that INTERPOS-BF demonstrates considerable improvement by 43.57\% over \textit{FEARec}, while INTERPOS-GF and INTERPOS-MF demonstrate 41.06\% and 36.22\% improvements over the best baseline.
Similarly, a  27.19\% and 37.97\% improvement is shown by INTERPOS-BF on \textit{HIT@15} and \textit{HIT@20} evaluation metrics over \textit{TiSASRec}.

\begin{table*}[t!]
\centering
\caption{Results on {\mbr} dataset, best results are shown in bold, second best results are underlined}
\vspace{-4pt}
%\resizebox{17cm}{!}{
\begin{tabular}
{l|cc|ccc|ccc}
\toprule
\multicolumn{1}{c|} {\multirow{2}{*}{\textbf{Category}}} & \multicolumn{2}{c|} {\multirow{2}{*}{\textbf{Method $\downarrow$ Metric $\xrightarrow{}$}}}                                                                    & \multicolumn{3}{|c|}{\textbf{NDCG}}                                                                   & \multicolumn{3}{c}{\textbf{HIT}}                                                                    \\ 
& &                                                                                 & \textbf{@10}                    & \textbf{@15}                    & \textbf{@20}                        & \textbf{@10}                    & \textbf{@15}                    & \textbf{@20}                    \\ \hline
\textbf{Popularity} & \multicolumn{2}{l|}{Pop} & 0.0077                           & 0.0092                           & 0.0103                           & 0.0151                           & 0.0208                           & 0.0256                           \\ \hline
{\multirow{7}{*}{\textbf{Sequential}}} & \multicolumn{2}{l|}{GRU4Rec~\cite{tan2016improved}}                                                                         & 0.0074                           & 0.0089                           & 0.0102                           & 0.0153                           & 0.021                            & 0.0261                 \\
& \multicolumn{2}{l|}{LightSANs~\cite{fan2021lighter}}                                                                       & 0.0079                           & 0.0092                           & 0.0104                           & 0.0158                           & 0.0208                           & 0.0260    \\
& \multicolumn{2}{l|}{SASRec~\cite{kang2018self}}                                                                          & 0.0087                           & 0.0103                           & 0.0116                           & 0.0172                           & 0.0233                           & 0.0288   \\
& \multicolumn{2}{l|}{SINE~\cite{tan2021sparse}}                                                                            & 0.0076                           & 0.0091                           & 0.0102                           & 0.0157                           & 0.0213                           & 0.0260  \\
& \multicolumn{2}{l|}{HGN~\cite{ma2019hierarchical}}                                                                             & 0.0046                           & 0.0056                           & 0.0064                           &0.0096                           & 0.0132                           & 0.0165 \\
& \multicolumn{2}{l|}{GCSAN~\cite{xu2019graph}}                                                                           & 0.0081                           & 0.0095                           & 0.0107                           &0.0161                           & 0.0214                           & 0.0266    \\
& \multicolumn{2}{l|}{BERT4Rec~\cite{sun2019bert4rec}}                                                                        & 0.0069                           & 0.0081                           & 0.0091                           &  0.0139                           & 0.0185                           & 0.0226    \\ \hline
{\multirow{2}{*}{\textbf{Time-aware}}} & \multicolumn{2}{l|}{FEARec~\cite{du2023frequency}} &0.0077 & 0.0089 & 0.0071 & 0.0169 & 0.0221 & 0.0231\\
\textbf{} & \multicolumn{2}{l|}{TiSASRec~\cite{li2020time}} & 0.0063 & 0.0091 & 0.0107 & 0.0167 & 0.0219 & 0.0259  \\ \hline
\multicolumn{1}{c|}{\multirow{6}{*}{\textbf{This work}}} &
\multicolumn{1}{c|}{\multirow{3}{*}{\textbf{LightSANs}}} & \multicolumn{1}{c|}{\textbf{INTERPOS-BF}} & 0.0083                           & 0.0098                           & 0.0111                           & 0.0167                           & 0.0224                           & 0.0279   \\
& \multicolumn{1}{c|}{}                                    & \multicolumn{1}{c|}{\textbf{INTERPOS-GF}} & 0.0091                           & 0.0107                           & 0.0120                           & 0.0186                           & 0.0246                           & 0.0301                           \\
& \multicolumn{1}{c|}{}                                    & \multicolumn{1}{c|}{\textbf{INTERPOS-MF}} & 0.0082                           & 0.0097                           & 0.0110                           & 0.0167                           & 0.0224                           & 0.0278                    \\ \cline{2-9}
& \multicolumn{1}{c|}{\multirow{3}{*}{\textbf{SASRec}}}    & \multicolumn{1}{c|}{\textbf{INTERPOS-BF}} & 0.0089                           & 0.0108                           & \underline{0.0125}                           &0.0184                           & \underline{0.0256}                           & \textbf{0.0328} \\
& \multicolumn{1}{c|}{}                                    & \multicolumn{1}{c|}{\textbf{INTERPOS-GF}} & \textbf{0.0094}                           & \textbf{0.0111}                           & \textbf{0.0127}                           & \textbf{0.0193}                           & \textbf{0.0257}                           & \underline{0.0322}   \\
& \multicolumn{1}{c|}{}                                    & \multicolumn{1}{c|}{\textbf{INTERPOS-MF}} & \underline{0.0093}                           & \textbf{0.0111}                           & \underline{0.0125}                           & \underline{0.0191}                           & \underline{0.0256}                           & 0.0318                    \\
\bottomrule
\end{tabular}
\label{tbl:mbr_results}
\vspace{-10pt}
\end{table*}

\stitle{Performance on Puzzle dataset.}
On {\mbrpuzzle} dataset, INTERPOS-GF turns out to be the outstanding fusion strategy. 
In comparison with the most competitive baseline \textit{FEARec} on \textit{NDCG@10}, INTERPOS-GF shows 52.80\% improvement.
% In comparison with the best baseline \textit{LightSANs}~\cite{fan2021lighter}, on \textit{NDCG@5}.
Similarly, INTERPOS-GF obtains a performance improvement of 53.48\% and 56.61\% over the best-performing baseline (i.e., \textit{TiSASRec}) on \textit{NDCG@15} and \textit{NDCG@20}, respectively. 
INTERPOS-BF gets 42.25\% and INTERPOS-MF manages a 42.53\% improvement over \textit{TiSASRec} on \textit{NDCG@20}. %in as shown in Table ~\ref{tbl:puzzle}.
% on \textit{NDCG@10}, we observe a noticeable improvement of 78.17\% shown by INTERPOS-GF over the best baseline, \textit{LightSANs}~\cite{fan2021lighter} while INTERPOS-BF and INTERPOS-MF show an improvement of 64.63\% and 61.57\%. 
INTERPOS outclasses the best baseline on \textit{HIT@10} and \textit{HIT@20} by significant margins, where INTERPOS-GF shows the highest improvement among the INTERPOS variations, 51.73\% and 64.75\%, respectively.  
% Refer to Table ~\ref{tbl:puzzle} for a more detailed snapshot of the results.

\stitle{Performance on MobileRec dataset.}
On the {\mbr} dataset, we observe INTERPOS with SASRec settings are the best-performing models, INTERPOS-GF outperforms all of our variants on all metrics with the exception of \textit{HIT@20}. 
%and \textit{SASRec} performs best among the baselines. 
We believe the key reason behind the INTERPOS integrated with SASRec variants outperforming LightSANs integrated INTERPOS variants is that {\mbr} has the least percentage of user interactions on the same day as compared to other datasets (shown in Figure~\ref{fig:rhythm_emergence_trends}).
Please see Table ~\ref{tbl:mbr_results} for detailed results. 
In the {\mbr} dataset, there are 19 million user-app interactions, 0.7 million unique users, and more than 10k unique applications. 
There are 27.56 interactions per user on average, while on average, there are more than 1800 interactions per unique app. 
% INTERPOS-GF outclasses all the INTERPOS variants (INTERPOS-BF and INTERPOS-MF) along with the competing baselines. 
On \textit{NDCG@10}, \textit{NDCG@15}, and \textit{NDCG@20}, INTERPOS-GF demonstrates 8.04\%, 7.76\% and 9.48\% improvement compared to the best competing baseline \textit{SASRec}.
On \textit{HIT@10}, \textit{HIT@15}, and \textit{HIT@20}, INTERPOS varitants outclass the best-performing baseline \textit{SASRec} by 12.20\%, 10.30\%, and 13.88\%, respectively. 

\stitle{Summary of Results.}
In conclusion, we observe that often existing time-aware baselines, FEARec and TiSASRe, perform better than sequential baselines, which highlights the significance of incorporating time sensitivity into the models.
However, these time-aware baselines incorporated time sensitivity at a later stage in contrast to {\ourmodel}, which fuses this information earlier on and produces significantly better results. 
Empirical results show that there is no clear winner in the fusion strategies. Yet, all fusion strategies exhibit statistically significant performance compared to all the baselines including sequential and time-aware recommendation systems across all 7 datasets. 
Moreover, fusion strategies integrated into \textit{SASRec} demonstrate superior performance on {\mbr}, while these strategies show better results on all other datasets when incorporated into \textit{LightSANs}.
It is also important to highlight despite not being a clear winner in the fusion strategies, the results of these strategies are statistically insignificant when compared with with another.
We empirically show that INTERPOS can better capture the user's behavioral and preferential shifts over the \textit{Activity Window}.
% For example, compared to SASRec~\cite{kang2018self} on average across all the datasets, in the \textit{NDCG@5} metric, INTERPOS-BF shows 32.52\% improvement, INTERPOS-MF gets 47.71\% and INTERPOS-GF manages to get 50.94\% improvement.
% Similarly, strong average improvements across all the datasets are demonstrated by INTERPOS on \textit{NDCG@k} for all values of $k$. 
% Large average improvements are observed across all the datasets on \textit{HIT@K} for all values of $k$. 
% Keeping in view the empirical findings, we argue that the user's interaction rhythm encodes informative behavioral and preferential shifts of a user over the \textit{Activity Window}. 
Through these results, we demonstrate that the integration of the user's interaction rhythm into an autoregressive next-item prediction model can facilitate better learning of the user's behavioral pattern leading to tailored predictions.

\balance
\section{Related Work}
Recommender systems have been applied in diversified domains like products~\cite{yan2022personalized, tanjim2020attentive}, news~\cite{wu2020mind}, movies~\cite{diao2014jointly} and a large body of research work focuses on improving recommender systems, which is the focus of this work.
% Prior art is abundant in sequential recommender systems. 
SASRec~\cite{kang2018self} strives to find a balance between Markov Chain~(MC) based methods and Recurrent Neural Network (RNN) based designs. %Markov chain-based methods rely on Markov property that future states of the system rely on the current state. 
%RNN-based methods tend to incorporate long contextual dependencies.  SASRec~\cite{kang2018self} proposes a transformer-based design to frame the next-item prediction task in an attempt to strike a balance between these two design principles. 
BERT4Rec~\cite{sun2019bert4rec} proposed to adopt a cloze objective for randomly masked items prediction by leveraging their bidirectional contextual conditioning. 
~\cite{huang2018improving} proposes a knowledge-enhanced memory network using RNN with key-value memory networks. 
%While these works enhance recommendation systems through RNN and transformer architecture, we leverage a transformer-based architecture to improve recommendation systems by integrating interaction rhythm.

He and others~\cite{he2016fusing} fused similarity-based methods with Markov chains for personalized sequential recommendations.
Observing that session embeddings and item embeddings are not in the same embedding space, ~\cite{hou2022core} proposes a framework to unify the representation spaces for encoding and decoding processes. ~\cite{ren2019repeatnet} proposes a model to make an informed decision on the consumption of repeated items. 
A short-term attention memory priority model is proposed in~\cite{liu2018stamp} %for capturing the user's generic interests from the interaction context while maintaining the importance of the user's recent interests. 
~\cite{zhou2020s3} suggested extracting the self-supervision signal by utilizing the inherent correlation in the data to improve the data representation through pertaining. 
Convolutional filters are employed in~\cite{tang2018personalized} %to learn the sequential patterns by embedding the interaction sequence into an image. 
~\cite{hidasi2016parallel} proposes parallel RNNs to exploit user clicks and accompanying features (visual and textural) for modeling user interaction sessions. 
~\cite{ying2018sequential} proposes a two-layer hierarchical attention network to make use of the user's long-term historical interactions and short-term preferences. 
In comparison, our work fuses the user's interaction rhythm to improve sequential recommendation systems.

Yuan and others~\cite{yuan2019simple} presented the idea of a holed convolutional neural network. %without relying on pooling operations and also utilizes residual block structure in the recommender system design. 
~\cite{zhang2019feature} argues to employ explicit features on top of the transition patterns of items %and utilizes weighted integration of item features into the item sequence. 
Graph neural networks are employed in ~\cite{wu2019session} %to capture complex item transitions and model the session sequences into graphs. 
%The attention network is utilized for composing global preferences and the user's current interests. 
%We also use the attention mechanism in the context of the user's interaction rhythm, which is significantly different compared to prior work. 
% Tackling the challenge of modeling ``third-order'' interactions between the users, their previous interactions, and the next item to be predicted, ~\cite{he2017translation} proposes a unified method to model these "third-order" relationships. 
% ~\cite{wu2020sse} proposes personalized transformer model.
%Improving recommendation systems for cold users is another direction of research, 
A neural personalized embedding model is proposed in~\cite{nguyen2018npe} for improving the recommendation performance for cold users.
The authors in~\cite{liu2021augmenting} proposed to employ transformer pretraining on reverse sequences and obtain predictions on the prior items. 
%By utilizing the pre-trained transformer, generated sequences (with fabricated items at the start) are employed to fine-tune the transformer model. In comparison, we use interaction rhythm to enhance recommendation systems. 
~\cite{li2020time} considers the user's interaction timestamp within the sequence modeling. 
~\cite{aarsynth},~\cite{pprior},~\cite{mobileconvrec} study app reviews in context of developer responses and app issues.
%We incorporate timestamps of user interactions to derive interaction rhythm and finally fuse interaction rhythm to improve sequential recommendation systems. 
\section{Conclusion and Future Work}
In this work, we propose {\ourmodel} which employs a user's interaction rhythm to morph the position encodings to inject a sense of the user's behavior and preference shifts over the user's interactions in the context of mobile apps recommendation. 
We empirically establish that interaction rhythm is correlated with a user's behavioral pattern and can provide an autoregressive model with a unique perspective on a user's preference shifts. 
We incroporate three strategies, INTERPOS-BF, INTERPOS-GF, and INTERPOS-MF, into two transformer-based recommendation system architecture. 
We show that the proposed integration strategies show strong learning capacity and help improve the underlying autoregressive model for the mobile apps recommendation task. 
We empirically establish that INTERPOS variants outclass the sequential and time-aware baselines across all the datasets by a large margin.
% We present three strategies, INTERPOS-BF, INTERPOS-GF, and INTERPOS-MF, to incorporate the interaction rhythm into an autoregressive next-item prediction model. 
% We also integrate INTERPOS fusion variants into a popular next-item prediction model, SASRec, and empirically establish that, on average, across all the datasets, INTERPOS variants outclass the SASRec by a large margin. 
In the future, we will explore alternative strategies to incorporate the user's interaction rhythm into an autoregressive design. 

\bibliographystyle{IEEEtran}
\bibliography{IEEEabrv,src/references}

% \clearpage
% \appendix
% \input{Sections/j_appendix}
\end{document}